# Electromagnetic cloaking in the visible frequency range.

**I. I. Smolyaninov, Y. J. Hung, and C. C. Davis**

Department of Electrical and Computer Engineering, University of Maryland,

College Park, MD 20742, USA

Metamaterials provide unprecedented freedom and flexibility in the creation of new devices, which control electromagnetic wave propagation in very unusual ways. Very recently various theoretical designs for an electromagnetic cloak have been suggested, and an experimental realization of a cloak operating in a two-dimensional cylindrical geometry has been reported in the microwave frequency range. In this communication we report on the experimental realization of the two-dimensional dielectric permittivity distribution required for non-magnetic cloaking in the visible frequency range.

Metamaterials provide unprecedented freedom and flexibility in the creation of new devices, which control electromagnetic wave propagation in very unusual ways. Very recently various theoretical designs of an electromagnetic cloak have been suggested [1-4]. An experimental realization of the electromagnetic cloak in the microwave frequency range has been reported in a two-dimensional cylindrical geometry [5]. Even though electromagnetic cloaking cannot be perfect, and it may be achieved only in a narrow frequency range, there may be practical benefits of a reduced visibility in various applications, which explains high current research interest in this topic.

Unfortunately, the most popular design of an electromagnetic cloak based on conformal coordinate transformations could not be easily implemented in the visible frequency range. The main reason is the need to vary magnetic permeability of the metamaterials in a non-trivial way, which is difficult to implement at optical frequencies. However, there exists a design for a non-magnetic optical cloak, which has been suggested by Shalaev *et al.* [4] in which this difficulty has been alleviated for a specific polarization state of the illuminating light. In effect, in this approach the electromagnetic field may be treated as a scalar field, and the only metamaterial parameter necessary to control remains the radial component of the dielectric permittivity $\varepsilon_r$:

$$\varepsilon_r = \left(\frac{r_2}{r_2 - r_1}\right)^2 \left(\frac{r - r_1}{r}\right)^2 \qquad (1)$$

where $r_1$ and $r_2$ are the internal and external radii of the cloak. In this communication we report on the experimental realization of the dielectric permittivity distribution approximately described by Eq.(1) in the frequency range around 500 nm in a two-dimensional geometry. Our approach is based on the plasmonic metamaterials described in detail in refs. [6-8], which are ideally suited for experimental realization of the scalar electromagnetic cloak described in ref. [4], since the surface plasmon-polariton (SPP) field has only one polarization state [9].

Our plasmonic metamaterials are based on layers of polymethyl methacrylate (PMMA) deposited on a gold film surface. In the 500 nm frequency region PMMA exhibits effective negative refraction as perceived by surface plasmons (the group velocity is opposite to the phase velocity, Fig.1(a)) [6]. We have fabricated various surface patterns consisting of stripes of PMMA separated by uncoated regions containing gold/air interfaces. The local orientation of PMMA stripes may be either parallel as shown in Fig.1(b,c), or slanted as shown in Fig.2(c,d,f). The width of the PMMA stripes $d_2$, the width $d_1$ of the gold/air portions of the interface, and the relative angle of the stripes may be chosen freely. Figs.2(a,b) demonstrate that the developed metamaterial allows a high degree of control of SPP propagation: narrow plasmon rays may be formed as a result of focusing and repeated self-imaging of the focal spot by the negative index stripes, as reported in ref.[6]. The successive stripes of positive and negative refractive index may continuously redirect the plasmon ray propagation along some curvilinear path, as shown in Figs. 2(b,c,f). The ability to bend the light path around some given area constitutes a necessary condition for a successful cloaking experiment.

The internal structure of the 2D plasmonic cloak [10] is shown in Figs. 1(b,c) and 4(a). It consists of concentric PMMA rings deposited on a gold film surface, in which the width of the PMMA rings $d_2$ and the width $d_1$ of the gold/vacuum portions of the interface was varied during the E-beam fabrication process. Since gold/air portions of the interface have effective group index $n_1>0$, by changing $(d_1+d_2)$ and the $d_1/d_2$ ratio, the average group refractive index of the multilayer material

$$n_{av}=(n_1 d_1 + n_2 d_2)/(d_1+d_2) \qquad (2)$$

may be continuously varied locally from large negative to large positive values. The distribution of $n_{av}$ in the fabricated 2D cloak is shown in Fig.1(d). It is reasonably close to the theoretical distribution given by eq.(1). Our numerical simulations performed using COMSOL multiphysics 3.3a validate this approach. The effective refractive index of the variable-diameter negative index ring structure is made reasonably close to the distribution given by eq.(1), which produces satisfactory cloaking performance.

In our experiments the plasmonic cloaking structure was illuminated by an external laser operating at 532 nm at an illumination angle that provides phase-matching excitation of SPPs at the left and right outer rims of the structure. Since the periodicity of the structure changes away from the outer rim, SPPs are not excited anywhere else, and the picture of light scattering presented in Figs.4(b,c) corresponds to SPP propagation inside the structure. Fig. 4(b) demonstrates that most of the plasmon energy cannot penetrate beyond the internal radius $r_1$ of the cloak, which corresponds to the $n_1 d_1 = - n_2 d_2$ (or $\varepsilon_r=0$) boundary.

However, a very small fraction of the plasmon rays, which propagate exactly through the center of the cloak, do reach the "cloaked" circular PMMA region in the middle of the structure. This practical difficulty has been mentioned in [1] as an unavoidable singularity, since these rays do not know whether to deviate left or right. On the other hand, compared to the circular PMMA region shown in the inset, which is surrounded by a concentric ring structure not optimized for cloaking, the amount of scattered energy has been reduced considerably. Fig.4(c) demonstrates the flow of plasmon energy around the cloaked region. Visualization is achieved due to weak scattering of plasmons into photons by the edges of the PMMA rings, which is observed using a regular optical microscope. Since scattering efficiency depends on the orientation of these edges, the optical signal appears to be considerably weaker in the top and bottom of the image (the image in Fig.4(c) was obtained by considerable overexposure, compared to the image in Fig.4(b)).

Thus, the basic properties of the electromagnetic cloak: i) considerable isolation of the cloaked region, and ii) the flow of energy around the cloak boundary, appear to match theoretical predictions. To our knowledge, this is the first experimental demonstration of the electromagnetic cloaking in the visible frequency range.

This work was supported in part by NSF grants ECS-0304046 and CCF-0508213.

**Figure Caption.**

Figure 1: (a) Real and imaginary parts of the wavevector of the symmetric surface plasmon mode propagating along 50 nm thick gold film in the PMMA/gold/glass and vacuum/gold/glass geometries as a function of frequency. In the frequency range marked by the box PMMA areas have negative refractive index as perceived by plasmons, while gold/vacuum interface looks like a medium with positive refractive index. The antisymmetric plasmon mode exhibits very high propagation losses and is not shown. (b) and (c) show AFM images of the central area of the 2D cloak at different zooms. (d) The distribution of $n_{av}$ in the fabricated 2D cloak compared to the theoretical distribution given by eq.(1).

Figure 2: (a) Plasmon ray propagation in a "magnifying superlens" concentric ring structure from ref.[6]. (b) Bending of the plasmon ray by the slanted array of PMMA stripes

in a parabolic lens structure shown in (d). Ray optics simulation of beam bending by a stack of slanted negative index layers is shown in (c). The refractive index of grey stripes is assumed to be $n_2= -1$. (e) and (f) show numerical simulations of the same effect performed using COMSOL multiphysics 3.3a.

Figure 3: (a) Numerical validation of the 2D cloaking design based on the variable diameter negative index rings performed using COMSOL multiphysics 3.3a. Distribution of the magnetic field is shown upon the illumination of the cloaking structure from the left. (b) Refractive index distribution within the 2D cloaking structure shown in (a).

Figure 4: (a) Two plasmonic cloak structures are observed using an optical microscope with white light illumination. The inset shows an AFM image of the central area of the cloak. (b) Optical image of surface plasmon-polariton propagation through these structures at 532 nm. The area inside the circle of radius $r_1$ is cloaked, except for a very small fraction of plasmon rays, which propagate exactly through the centre of the cloak. The illumination direction is shown by the arrow. The inset demonstrates plasmon scattering by a typical concentric ring structure, which is not optimized for cloaking at 532 nm. The plasmons are strongly scattered by the edge of the structure and by the circular area in the middle. (c) False color representation of the measured plasmon field scattering around the central area of the cloak. The flow of energy around the cloaked region is visualized.

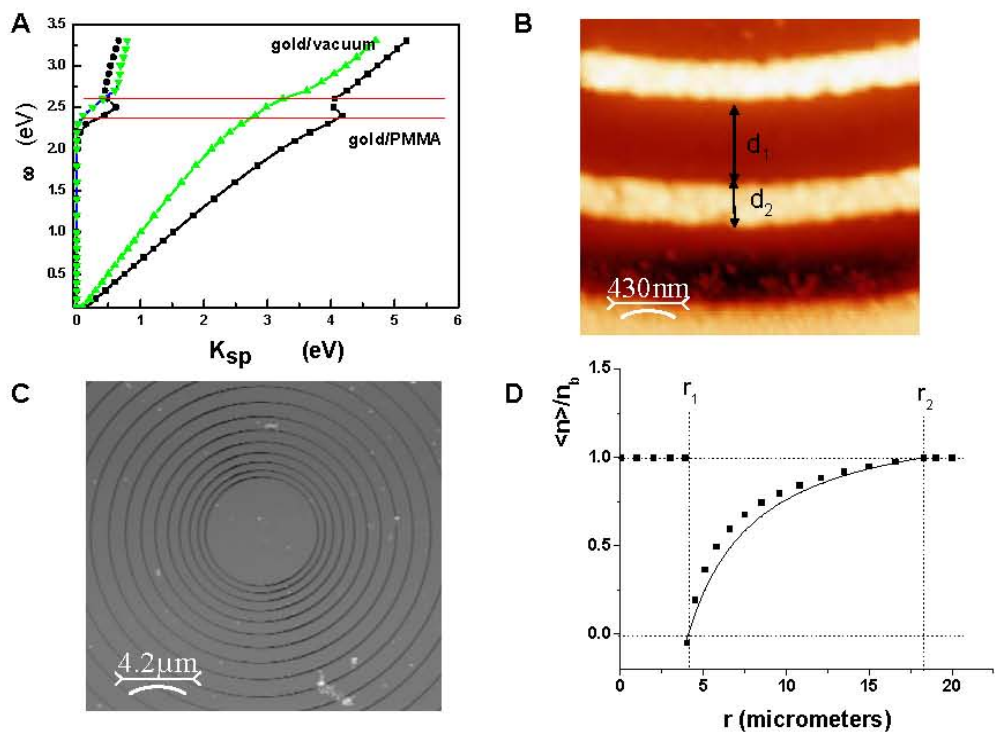

Fig.1

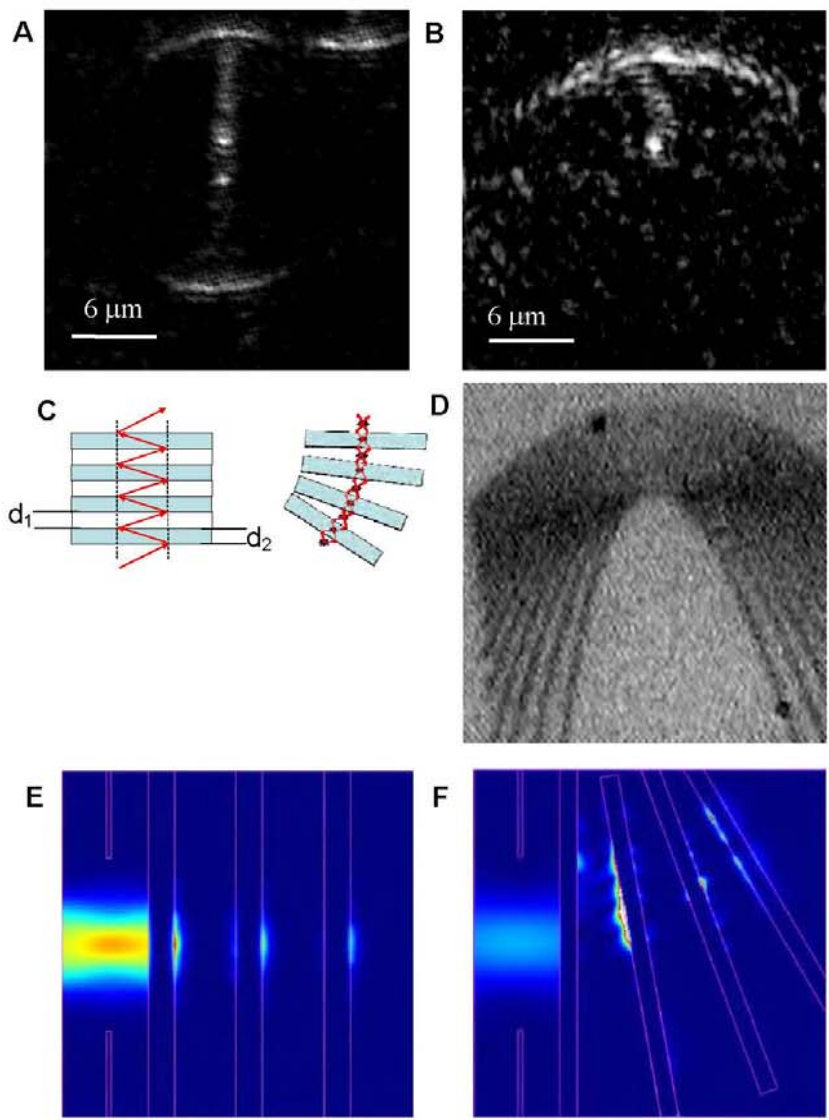

Fig.2

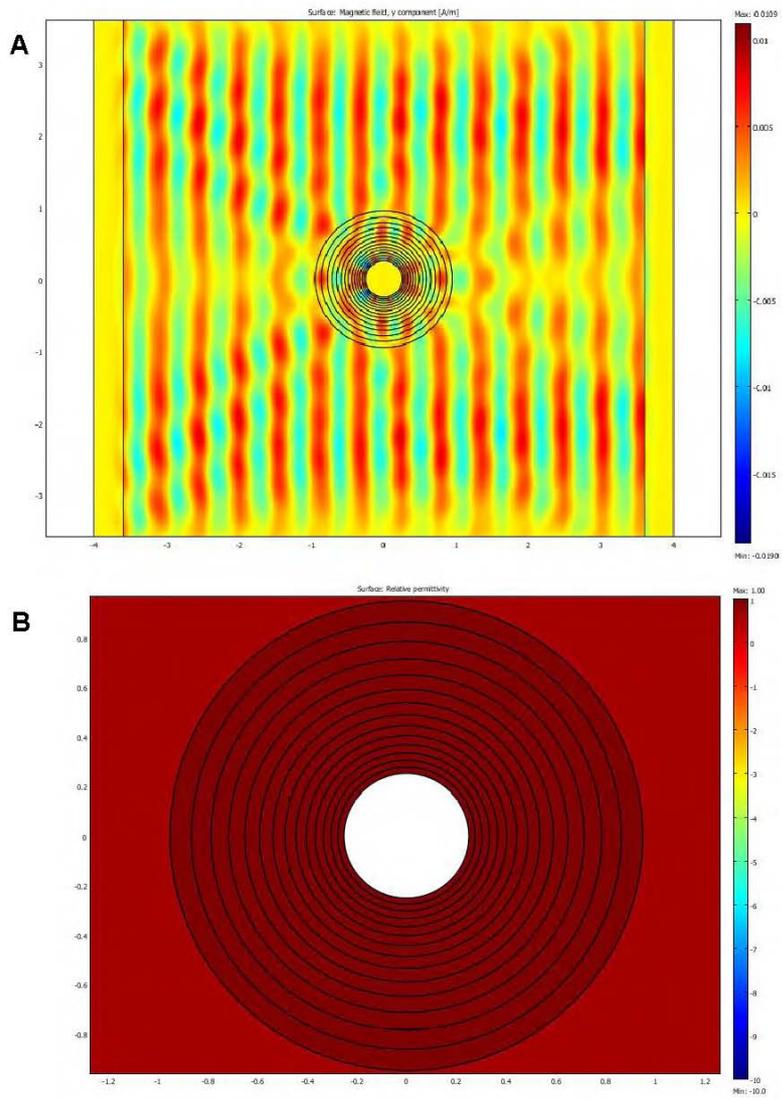

Fig.3

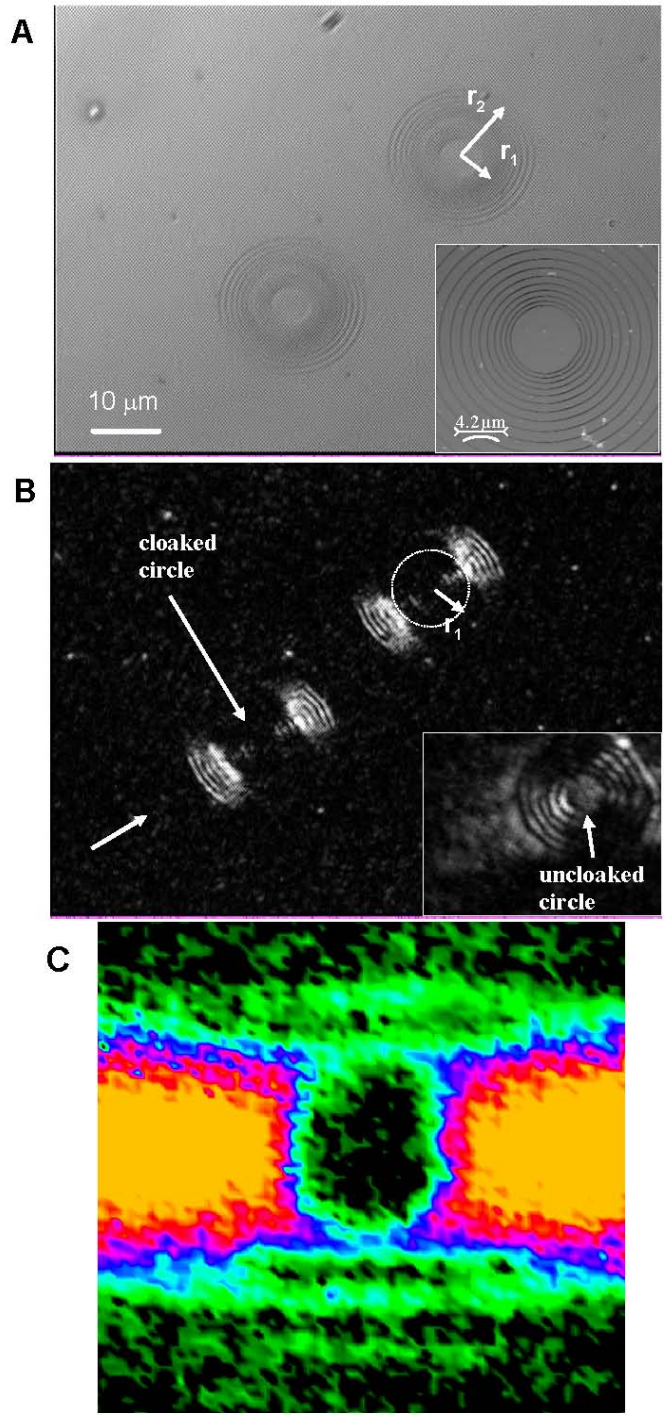

Fig.4